\documentclass[aps, pra, onecolumn,notitlepage, amsmath, amssymb, superscriptaddress, nofootinbib]{revtex4-1}

\makeatletter
\def\p@subsection{}
\def\p@subsubsection{}
\makeatother

\usepackage[overload]{empheq}
\usepackage{cases}
\usepackage[utf8]{inputenc}
\usepackage{graphicx}
\usepackage{units}
\usepackage{color}
\usepackage[pdftex, colorlinks=true, linkcolor=myblue, citecolor=myblue,
urlcolor=myblue]{hyperref}
\usepackage{times}
\usepackage{comment}
\usepackage{graphicx}
\usepackage{amsmath, calc}
\usepackage{mathrsfs}
\usepackage{MnSymbol}
\usepackage{amsfonts}
\usepackage{amssymb}
\usepackage{bm}
\usepackage{bbm}
\usepackage{color}
\usepackage{array}
\usepackage{units}
\usepackage{mathrsfs,amsmath}
\usepackage{mathtools}
\definecolor{myblue}{rgb}{0,0,1}
\definecolor{myred}{rgb}{1,0,0}
\usepackage{float}
\usepackage{cancel}

\newcommand{\ket}[1]{|#1\rangle}

\newcommand{\kettwo}[1]{||#1\rangle \rangle}

\newcommand{\da}{\dagger}

\begin{document}
\title{Doublons, topology and interactions in a one-dimensional lattice}

\author{P. Mart\'{i}nez Azcona}
\affiliation{Departamento de F\'{i}sica de la Materia Condensada, Universidad de Zaragoza, Zaragoza 50009, Spain}

\author{C. A. Downing} 
\email{c.a.downing@exeter.ac.uk}
\affiliation{Department of Physics and Astronomy, University of Exeter, Exeter EX4 4QL, United Kingdom}

\date{\today}

\begin{abstract}
We investigate theoretically the Bose-Hubbard version of the celebrated Su-Schrieffer-Heeger topological model, which essentially describes a one-dimensional dimerized array of coupled oscillators with on-site interactions. We study the physics arising from the whole gamut of possible dimerizations of the chain, including both the weakly and the strongly dimerized limiting cases. Focusing on two-excitation subspace, we systematically uncover and characterize the different types of states which may emerge due to the competition between the inter-oscillator couplings, the intrinsic topology of the lattice, and the strength of the on-site interactions. In particular, we discuss the formation of scattering bands full of extended states, bound bands full of two-particle pairs (including so-called `doublons', when the pair occupies the same lattice site), and different flavors of topological edge states. The features we describe may be realized in a plethora of systems, including nanoscale architectures such as photonic cavities and optical lattices, and provide perspectives for topological many-body physics.  
\end{abstract}


\maketitle



\section{Introduction}
\label{Intro}

The field of topological nanophotonics investigates how to exploit the geometric and topological properties of photonic systems in order to design and control both classical and quantum light~\cite{Lu2014, Ozawa2019, Smirnova2020}. The field continues to display some fascinating physics, from unidirectional propagation of optical states immune to disorder~\cite{Wang2009}, to the development of topological lasers~\cite{Bandres2018}, to the creation of topologically-protected quantum light~\cite{Mittal2018, Blanco2018}. The simplicity and beauty of the field has ensured that many theoretical works do not need to stray beyond the single-particle level in order to describe and predict some fascinating topological photonic effects~\cite{Harari2018, Zhirihin2019, Savelev2020, Bobylev2021}. This is especially true when considering phenomena in photonic lattices inspired by iconic topological theories, such as the Harper-Hofstadter~\cite{Hofstadter1976}, Su-Schrieffer-Heeger~\cite{Su1979} and Haldane models~\cite{Haldane1988}.

Perhaps the simplest theory providing insight into the role of interactions in lattice models is the one-dimensional Bose-Hubbard model~\cite{Essler2005}. For the past few decades, low-dimensional interacting bosonic systems have been intensively studied with the aid of such models in order to describe systems like ultracold atoms in optical lattices~\cite{Lewenstein2007, Cazalilla2011, Krutitsky2016}. Pioneering experiments have revealed novel features due to particle-particle interactions, such as exotic bound states of pairs of ultracold atoms in a regularly spaced chain of microtraps~\cite{Winkler2006}. Since the Su-Schrieffer-Heeger model~\cite{Su1979} exhibits some interesting topological features with essentially just a dimerized (rather than regularly spaced) chain, it is is natural to ponder the interplay between interactions and topology in an extended Su-Schrieffer-Heeger-like model. Furthermore, as the first step towards understanding topological many-body systems, one may judiciously focus on the simplest nontrivial subspace with interactions, that of only two excitations. Indeed, the two-excitation sector already provides some hallmarks of multi-excitation physics, such as bound two-particle states and novel bands in the quasiparticle bandstructure~\cite{GorlachPoddubny2017, Salerno2018, Zurita2020, Pelegri2020}.

\begin{figure}[tb]
 \includegraphics[width=0.50\linewidth]{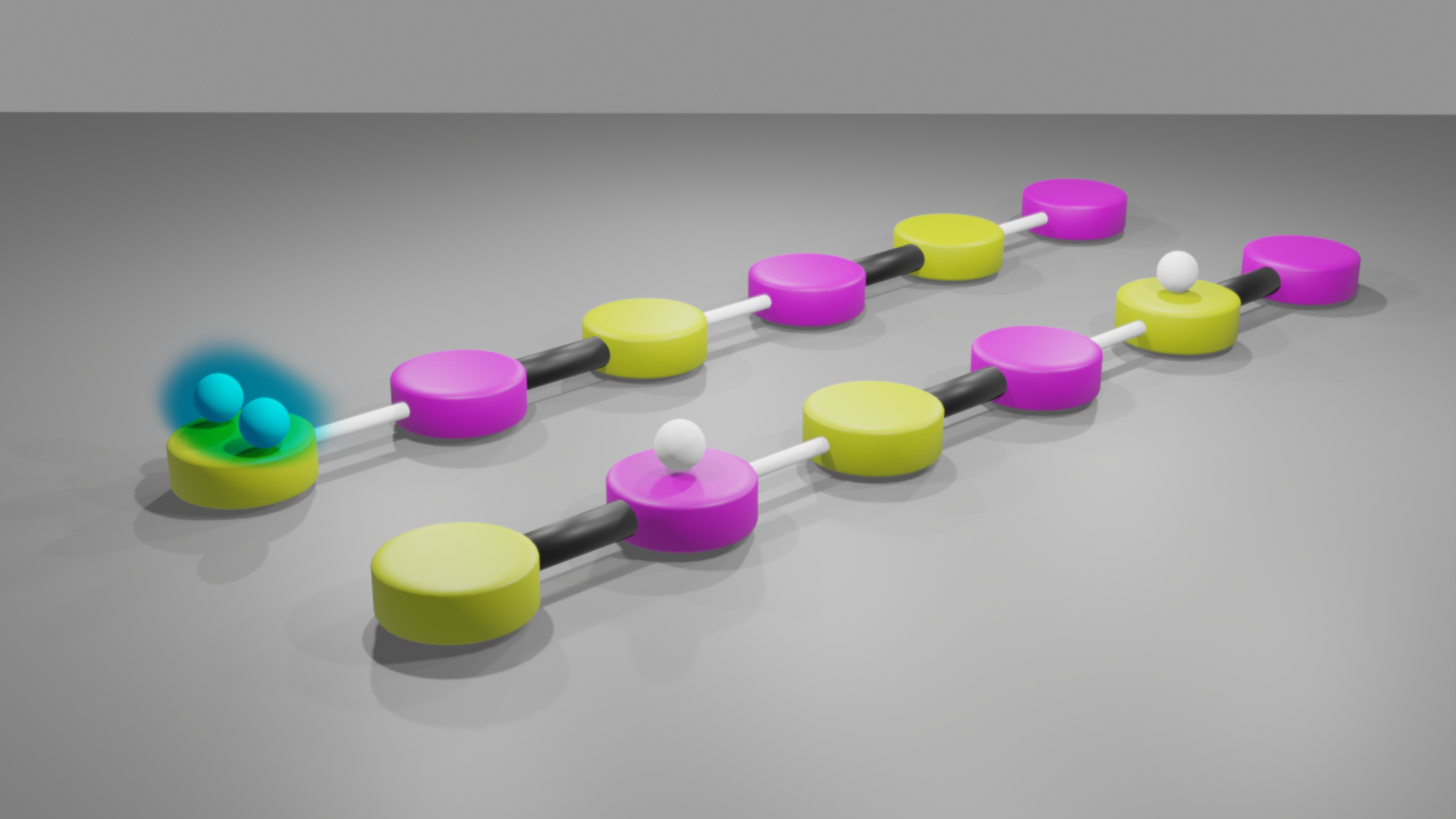}
 \caption{ Sketches of dimerized chains of oscillators, each of resonance frequency $\omega_0$, and represented by yellow and pink disks. The alternating coupling strengths $J_1$ and $J_2$ are denoted by thin white and thick black rods. The setup mimics the two-particle Bose-Hubbard version of the celebrated Su-Schrieffer-Heeger topological model. The two excitations are depicted by two white balls, and are subject to the on-site interaction $U$. Left chain: in the nearest edge oscillator there is a doublon (illustrated by the blue cloud) of two bound excitations on the same oscillator. Right chain: the two excitations are separated, and may be found in the bulk of the chain on the second and fifth oscillators respectively. }
 \label{SSH_System}
\end{figure}

Here we consider a dimerized chain of oscillators (each of resonance frequency $\omega_0$), characterized by the alternating coupling strengths along the chain $J_1 \ge 0$ and $J_2 \ge 0$, as sketched in Fig.~\ref{SSH_System}. The chain is then formed of $N$ dimers ($2N$ oscillators), and we suppose that any excitations are subject to an on-site interaction $U$. Therefore, the associated Hamiltonian $\hat{H}$ acts as a Bose-Hubbard version of the celebrated Su-Schrieffer-Heeger topological model~\cite{Su1979}, and reads~\cite{Gorlach2017, Liberto2016, Liberto2017, Recati2018, Marques2018, Stepanenko2020, Lyubarov2020, Tschernig2021, Stepanenko2021, SuppInfo}
\begin{equation}
\label{SSH_Ham_1}
\hat{H} = \: \omega_0 \sum_{n=1}^N \left( a^\da_n a_n + b^\da_n b_n \right)
 - J_1 \sum_{n=1}^{N} \left( b^\da_n a_{n} + a^\da_{n} b_n \right) - J_2 \sum_{n=1}^{N-1} \left( b^\da_n a_{n+1} + a^\da_{n+1} b_n \right)
 + \frac{U}{2} \sum_{n=1}^{N}  \left( a^\da_n a^\da_n a_{n} a_{n} + b^\da_n b^\da_n b_{n} b_{n} \right),
\end{equation}
where $a^\da_n$ and $b^\da_n$ ($a_n$ and $b_n$) create (destroy) a bosonic excitation on the $n$-th dimer in the chain, which is composed of $a$ and $b$ type oscillators (as signified by the yellow and pink disks respectively in the sketch of Fig.~\ref{SSH_System}). There are two key quantities in this nearest-neighbor tight-binding model: (i) the frequency scale $\bar{J}$, and (ii) the dimerization parameter $\epsilon$, which are defined using $J_1$ and $J_2$ by
\begin{equation}
\label{SSH_2}
\bar{J} = J_1 + J_2, \quad\quad\quad\quad\quad\quad \epsilon = \frac{J_1 - J_2}{\bar{J}}.
\end{equation}
Then the dimensionless ratio $U/\bar{J}$ measures the influence of the on-site interactions, while the dimerization parameter $\epsilon$ is purely geometric and tracks the influence of the intrinsic topology of the dimerized chain. 

\begin{figure}[tb]
 \includegraphics[width=0.50\linewidth]{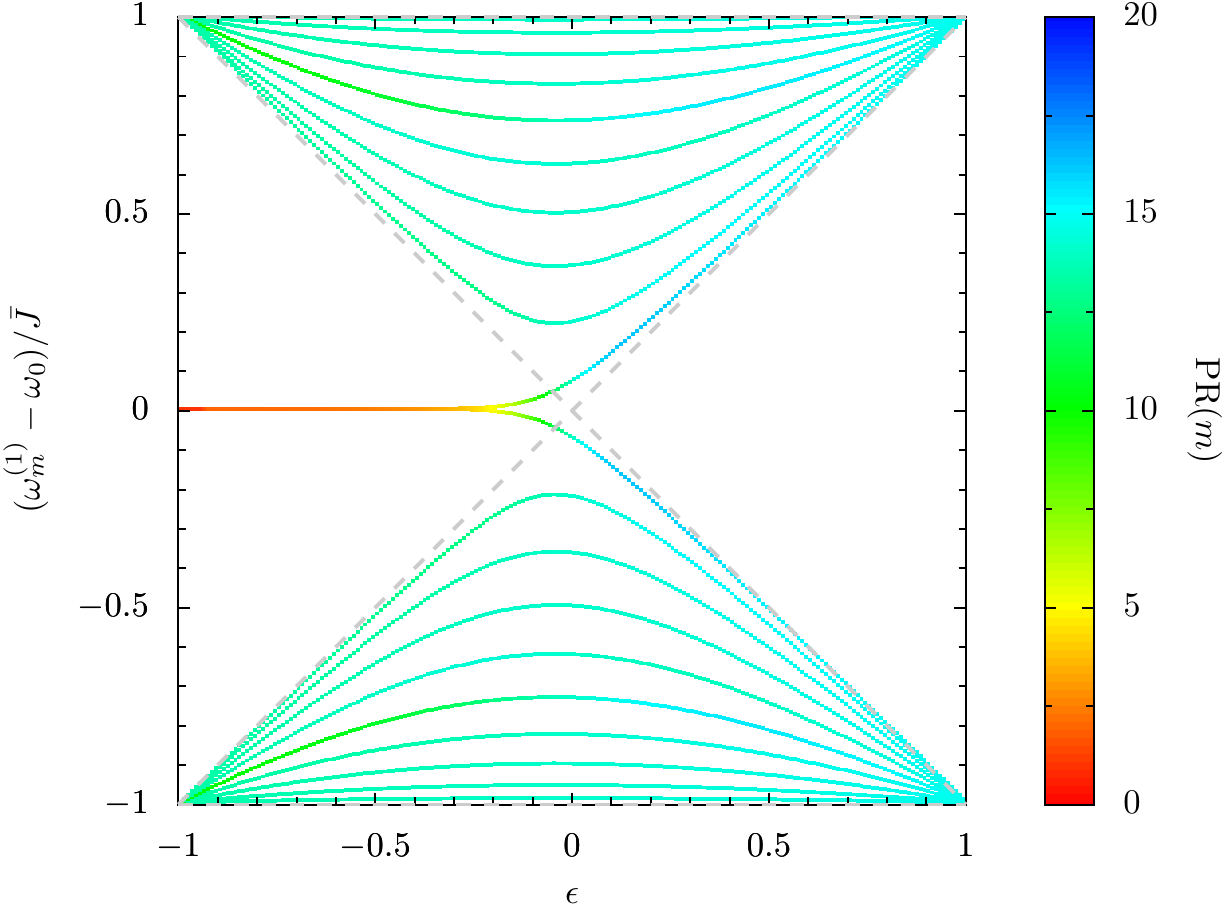}
 \caption{ Eigenfrequencies $\omega_{m}^{(1)}$ of a finite dimerized chain in the one excitation sector (as measured from $\omega_0$), in units of the coupling strength $\bar{J}$, as a function of the dimerization parameter $\epsilon$ [cf. Eq.~\eqref{SSH_2}]. Color bar: the participation ratio $\mathrm{PR}(m)$ of each state $m$~\cite{SuppInfo}. Edge states are highly localized and thus appear reddish on this scale. Dashed gray lines: region enclosing the extended states, as defined in the infinite chain limit~\cite{SuppInfo}. In the figure, the chain is composed of $N = 10$ dimers, leading to a $20$-dimensional Hilbert space.}
 \label{SSHdim}
\end{figure}

In the single-excitation sector, the interactions described by the final term in Eq.~\eqref{SSH_Ham_1} may be disregarded, and the model collapses into a simple one-particle Su-Schrieffer-Heeger-like topological model~\cite{Su1979, DowningWeick2017, DowningWeick2018}. In this regime, the Zak phase acts as the relevant topological invariant governing the system~\cite{Zak1989,SuppInfo}. It predicts a topologically trivial phase when $\epsilon \ge 0$ and the Zak phase is $0$, and a topologically nontrivial phase when $\epsilon < 0$ and the Zak phase is $\pi$~\cite{Asboth2016, SuppInfo}. Remarkably, the topologically nontrivial phase is associated with the presence of highly localized edge states, which reside at the mid-gap frequency $\omega_0$, between an upper band and a lower band full of highly extended states. The evolution of the system from topologically nontrivial to topologically trivial is shown for a finite system in Fig.~\ref{SSHdim}, as found by diagonalizing the $2N \times 2N$ matrix Hamiltonian following from Eq.~\eqref{SSH_Ham_1} in the single-excitation sector~\cite{SuppInfo}. We consider $N = 10$ dimers ($2N = 20$ oscillators), and we plot the single-excitation eigenfrequencies $\omega_{m}^{(1)}$ as a function of the dimerization parameter $\epsilon$. The color bar records $\mathrm{PR}(m)$, the participation ratio~\cite{SuppInfo, Bell1970, Thouless1974} of the eigenstate associated with the eigenvalue $\omega_{m}^{(1)}$. This quantity provides a basic measure of the degree of localization of the state, with a lower $\mathrm{PR}(m)$ corresponding to a more localized state, and a higher $\mathrm{PR}(m)$ corresponding to a more extended state, with $1 \le \mathrm{PR}(m) \le 2N$~\cite{note1}. The famous edge states shown at $\omega_0$ in Fig.~\ref{SSHdim} (red-orange in $\mathrm{PR}(m)$ on this color scale), have an intuitive geometric origin. When $\epsilon <0$ (or $J_2 > J_1$), it is energetically favorable for each oscillator in the chain to pair up with its neighboring oscillator with which it shares a strong $J_2$ link. This dimerization leaves the first and last oscillators unpaired, since they only connect to their nearest oscillator via a weak $J_1$ link, and they instead house edge states at the bare resonance frequency $\omega_0$. When $\epsilon \ge 0$, each oscillator in the chain pairs up with the neighboring oscillator with which it shares a strong $J_1$ link, leaving no oscillators unpaired throughout the chain and thus removing the opportunity for edge states to arise. In this rudimentary way, these bosonic edges states emerging from dimerized chains are the cousins of the celebrated Majorana zero-modes found in Kitaev chains~\cite{Kitaev2001}.

The two-excitation sector of Eq.~\eqref{SSH_Ham_1} presents a much richer topological structure due to the influence of the final interaction term, as governed by $U$, and it has been recently studied by several authors~\cite{Liberto2016, Liberto2017, Gorlach2017, Recati2018, Marques2018, Stepanenko2020, Lyubarov2020, Tschernig2021, Stepanenko2021}. In particular, Eq.~\eqref{SSH_Ham_1} leads to \textit{scattering states}, which are composed of superpositions of two-particle states not residing on the same oscillator site; and \textit{bound states}, which are made up of superpositions of states where the two excitations appear on the same oscillator site (so-called \textit{doublons}) or on the same dimer of oscillators. Furthermore, as in the one-particle regime represented in Fig.~\ref{SSHdim}, there are \textit{bulk states} (those extended throughout the chain) and \textit{edge states} (those pinned exponentially at the edges) due to the inherent topology of the array. Therefore, one may envisage that in the two-particle regime there is already a rather complicated zoo of states to classify and understand, due to the interplay of interactions and topology. In the remainder of this work we study in detail the properties of the two-particle Su-Schrieffer-Heeger model with interactions, at the simplest level of the prospective bandstructures and eigenstates. Importantly, we consider the whole range of possible dimerizations $\epsilon$ of the chain, including both the weakly dimerized and the strongly dimerized limits, and thus generalize Fig.~\ref{SSHdim} to capture the two-excitation regime.


\section{Results}
\label{doublons_section}

\begin{figure}[tb]
 \includegraphics[width=1.0\linewidth]{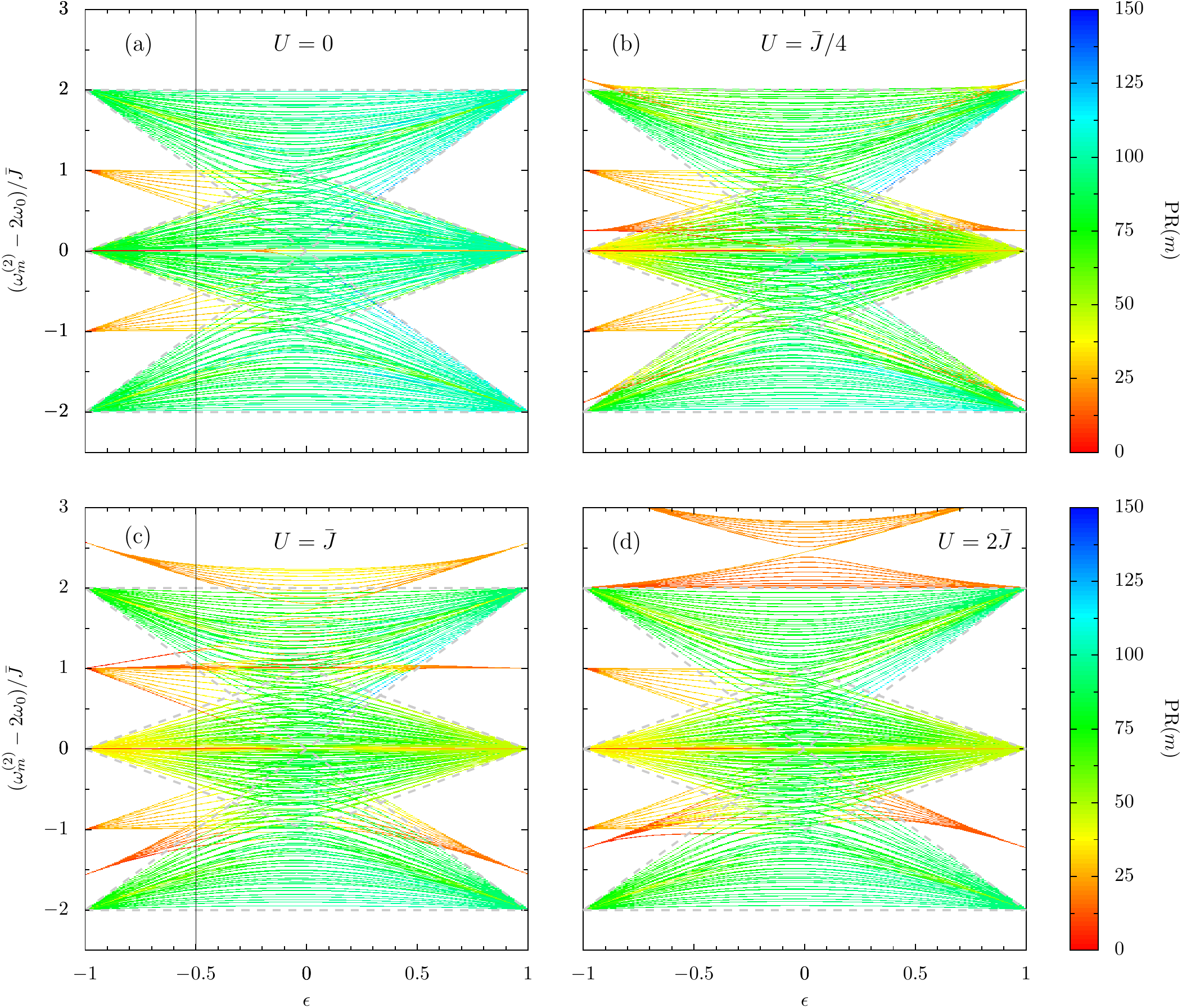}
 \caption{ Eigenfrequencies $\omega_{m}^{(2)}$ of a dimerized chain in the two excitation sector (as measured from $2\omega_0$, and in units of the coupling strength $\bar{J}$) as a function of the dimerization parameter $\epsilon$ [cf. Eq.~\eqref{SSH_2}]. The interaction strength $U$ increases upon going along the panels. Color bar: the participation ratio $\mathrm{PR}(m)$ of each state $m$~\cite{SuppInfo}. Highly localized states are red-orange in color, while extended state are blue-green. Dashed gray lines: regions enclosing the bulk scattering states, as calculated in the infinite chain limit~\cite{SuppInfo}. Solid gray vertical lines in panels (a) and (c): cuts at $\epsilon = -1/2$, as guides for the eye. In the figure, the chain is composed of $N = 10$ dimers, leading to a $210$-dimensional Hilbert space. }
 \label{diagSSH}
\end{figure}

The Hamiltonian of Eq.~\eqref{SSH_Ham_1} may be straightforwardly diagonalized numerically in the two-excitation sector, where a chain of $N$ dimers ($2N$ oscillators) necessitates working with a $N(2N+1) \times N(2N+1)$ matrix. This dimensionality arises because the basis contains $2N$ double-occupancy states, like $\kettwo{2, 0, 0,...}$, $\kettwo{0, 2, 0, ...}$ and so on, and $N(2N-1)$ single-occupancy states, like $\kettwo{1, 1, 0,...}$, $\kettwo{1, 0, 1, ...}$ and so on. Here we use $\kettwo{i, j, k, ...}$, which has $2N$ elements, to refer to bosonic eigenstates in the occupation number basis in the two-excitation sector. The two-particle eigenfrequencies $\omega_{m}^{(2)}$ (here $m = 1,2, ... N(2N+1)$ labels the state) are naturally measured from $2\omega_0$, while the frequency unit $\bar{J}$ is given by Eq.~\eqref{SSH_2}. The topology of the system can then be probed by changing the dimerization parameter $\epsilon$ [cf. Eq.~\eqref{SSH_2}]. We present the eigenfrequencies $\omega_{m}^{(2)}$ as a function of the full range of potential dimerizations $-1 \le \epsilon \le 1$ in Fig.~\ref{diagSSH} for a chain of $N=10$ dimers. The interaction strength is successively increased along the panels (a-d), with $U/\bar{J} = \{ 0, 1/4, 1, 2\}$. The dashed gray lines in all panels of Fig.~\ref{diagSSH} define regions enclosing the scattering states, as calculated in the infinite chain limit~\cite{SuppInfo}, which are eigenstates wholly comprised of single-occupancy states.

In Fig.~\ref{diagSSH}~(a) the simplest case is presented, that with vanishingly small interactions ($U \to 0$), such that the two-particle solutions are simple combinations of the single-particle solutions (which can themselves be classified as either bulk states or edge states~\cite{Asboth2016, SuppInfo}). Therefore, due to the necessary pairing up of single-particle results, one should find three categories of scattering state in the two excitation sector with $U \to 0$: bulk-bulk states, bulk-edge states, and edge-edge states. In the topologically trivial phase ($\epsilon \ge0$) three distinct scattering bands of bulk-bulk states (mostly blue-green in $\mathrm{PR}(m)$ on this color scale) can be seen, as neatly predicted by the continuum calculation (dashed gray lines)~\cite{SuppInfo}. In the strongly dimerized limit ($\epsilon \to 1$, such that $J_2 \to 0$) the bulk-bulk states amalgamate at the three eigenfrequencies $2\omega_0 \pm 2\bar{J}$ and $2\omega_0$, since the two constituent single-particle states contribute $\omega_0 \pm J_1$ (which are the two eigenfrequencies of a dimer coupled by $J_1$, in the single-excitation sector). In the opposing topologically nontrivial phase ($\epsilon < 0$), in addition to the three scattering bands familiar from $\epsilon \ge0$ are two scattering bands housing more localized bulk-edge states [mostly orange in $\mathrm{PR}(m)$]. The strongly dimerized limit ($\epsilon \to -1$, such that $J_1 \to 0$) reveals that the bulk-edge states merge at the two eigenfrequencies $2\omega_0 \pm \bar{J}$, since the elemental single-particle states contribute $\omega_0$ (from the edge states) and $\omega_0 \pm J_2$ (the two eigenfrequencies of a dimer coupled by $J_2$). The third class of state, the topologically protected edge-edge states, are found at $2 \omega_0$ (mostly red) and their existence may be envisaged from the single-particle physics displayed at $\omega_0$ in Fig.~\ref{SSHdim}.

Interactions are turned on in the remaining panels of Fig.~\ref{diagSSH}, from $U = \bar{J}/4$ to $U = \bar{J}$ to $U = 2\bar{J}$ in panels (b, c, d). Let us first consider the $\epsilon \ge 0$ regime, and for simplicity the highly dimerized limit $\epsilon \to 1$ (such that $J_2 \to 0$). Most noticeably, the effect of nonzero interactions in panels (b-d) is to introduce three new bound state bands (mostly orange in $\mathrm{PR}(m)$ on this color scale), which complement the three scattering bands (mostly blue-green) familiar from panel (a). With weaker interactions in panel (b), these bound state bands mostly overlap the scattering bands, since the energy cost $U$ is minimal compared to the coupling parameter $\bar{J}$. However, in panels (c) and (d) the increasingly strong interactions act to lift the frequencies of the bound state bands, such that in panel (d) two of the bound state bands naturally start to become detached from the upper scattering band. This bound state behavior can be captured exactly in this highly dimerized limit ($\epsilon \to 1$), where the chain collapses into $N$ disconnected Bose-Hubbard dimers, via the following expression (valid for a subset of $m$)~\cite{SuppInfo}
\begin{subequations}
\label{eq:po_0}
  \begin{align}[left ={ \lim_{\epsilon \rightarrow 1} \omega_{m}^{(2) } = \empheqlbrace}]
    & 2 \omega_0 + \frac{U}{2} \pm \sqrt{ \left( 2 \bar{J} \right)^2 + \left( \tfrac{U}{2} \right)^2 }, \label{eq:po_1} \\
    & 2 \omega_0 + U. \label{eq:po_2}
   \end{align}
  \end{subequations}
With $U \gtrsim \bar{J}$, the largest eigenfrequency [the expression with $+$ in Eq.~\eqref{eq:po_1}] is associated with the symmetric doublonic eigenstate $\left( \kettwo{20} + \kettwo{02} \right)/\sqrt{2}$. The intermediate eigenfrequency, as given by Eq.~\eqref{eq:po_2}, is independent of $J_1$ and $J_2$ and is associated with the antisymmetric doublonic eigenstate $\left( \kettwo{20} - \kettwo{02} \right)/\sqrt{2}$. Finally, the smallest eigenfrequency [the expression with $-$ in Eq.~\eqref{eq:po_1}] is associated with the third type of bound state, where the two excitations are equally shared amongst one specific dimer as $\kettwo{11}$ (that is, a bound state but not a doublon bound state). The physics of the $\epsilon < 0$ regime is again more complex due to the residual topology of the dimerized chain. In this case, there are five extra bands in Fig.~\ref{diagSSH}~(b, c, d) (where $U \ne 0$), as compared to panel~(a) (where $U \to 0$). This fact is most apparent in the the highly dimerized limit $\epsilon \to -1$, where, as well as the three bound state bands governed by Eq.~\eqref{eq:po_0} for $\epsilon \to 1$, there are two extra bands. These two additional bands arise due to the extra combinations of two-particle states which are possible because of the edge state-like and bulk state-like characters of the underlying single-particle states (namely, the remnants of the topological edge states well defined in the $U \to 0$ limit). 

\begin{figure*}[tb]
 \includegraphics[width=1.0\linewidth]{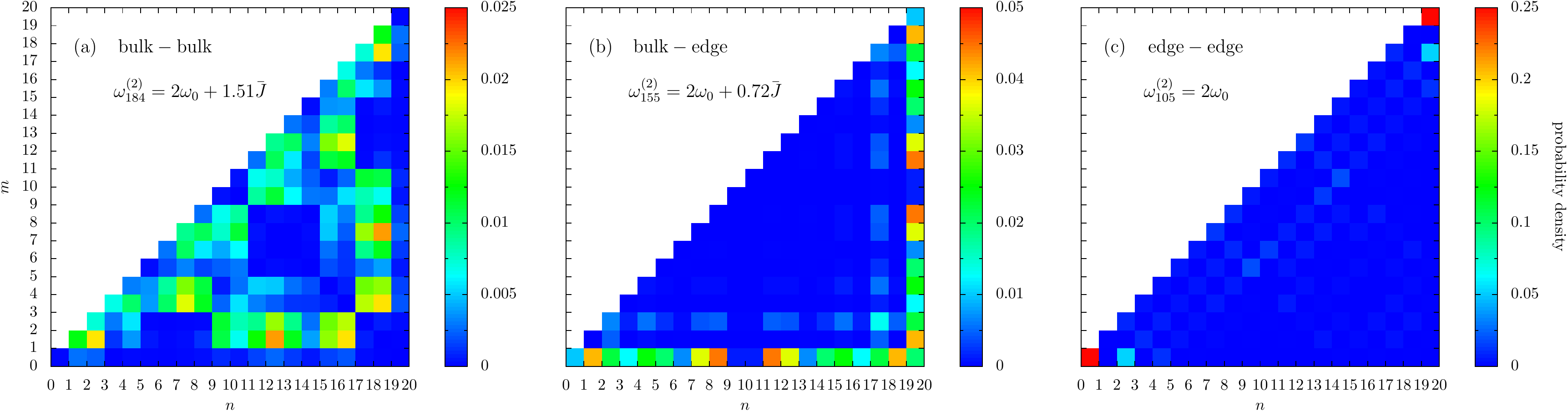}
 \caption{ Three typical eigenstates of a dimerized chain in the two excitation sector, with vanishing interactions ($U \to 0$). Color bars: probability densities (note: the scale is different for each panel). In the figure, the dimerization parameter $\epsilon = -1/2$ and the chain is composed of $N = 10$ dimers, leading to a $210$-dimensional Hilbert space. This regime corresponds to descending the thin, gray vertical line in Fig.~\ref{diagSSH}~(a). }
 \label{stateUzero}
\end{figure*}

The eigenstates of the dimerized chain in the two-excitation sector are associated with two integers $n$ and $m$, since each two-particle state can be decomposed (up to a normalization factor) by $c_n^\dagger c_m^\dagger \ket{\mathrm{vac}}$, where $\ket{\mathrm{vac}}$ is the (unoccupied) vacuum state and $c_n^\dagger = \{ a_n^\dagger, b_n^\dagger\}$ captures the two flavors of bosonic creation operator [cf. Eq.~\eqref{SSH_Ham_1}]. Using the two examples of possible state sketched in Fig.~\ref{SSH_System}, the doublon in the left chain is described by $a_1^\dagger a_1^\dagger \ket{\mathrm{vac}} / \sqrt{2}$, while the two excitations in the right chain are represented with $b_1^\dagger a_3^\dagger \ket{\mathrm{vac}}$. In this way, our considered two-excitation states live geometrically on a triangular eigenspace of base $n$ and height $m$ (after removing the redundant replication of information arising from the bosonic symmetry). Doublons occupy the same lattice site and so are codified by $n=m$, they are found along the hypotenuse of the triangle. Bound states are associated with either doublons or with singly-occupied states confined to the same dimer, hence they are found on the hypotenuse diagonal and the immediately adjacent squares. Scattering states can reside everywhere, and thus are exposed by their high participation ratios.

The solid gray vertical lines in panels (a) and (c) of Fig.~\ref{diagSSH} cut the plots at $\epsilon = -1/2$, a somewhat arbitrary point which is nevertheless characteristic of the more interesting $\epsilon <0$ regime, which we now investigate in more detail. In Fig.~\ref{stateUzero} we plot explicitly the eigenstates (where the interactions are vanishingly small, $U \to 0$) associated with three typical two-particle states highlighted in Fig.~\ref{diagSSH}~(a) by the predominantly green, orange and red participation ratios respectively. Namely, we show representative bulk-bulk states, bulk-edge states, and edge-edge states in panels (a), (b) and (c) respectively in Fig.~\ref{stateUzero}. The bulk-bulk state of Fig.~\ref{stateUzero}~(a) is characteristically spread throughout the triangular eigenspace, as are all extended states of a similarly high participation ratio. A bulk-edge state is represented in Fig.~\ref{stateUzero}~(b), which showcases a high probability density along the opposite and adjacent sides of the triangular eigenspace, due to the competition between the constituent single particle states (a topological edge state and a bulk state). In Fig.~\ref{stateUzero}~(c), one sees an example of a fully topological edge-edge state residing exactly at $2 \omega_0$, which clearly exhibits strong doublon-like localization at the ends of the chain on the bound state diagonal (red squares). 

\begin{figure*}[tb]
 \includegraphics[width=1.0\linewidth]{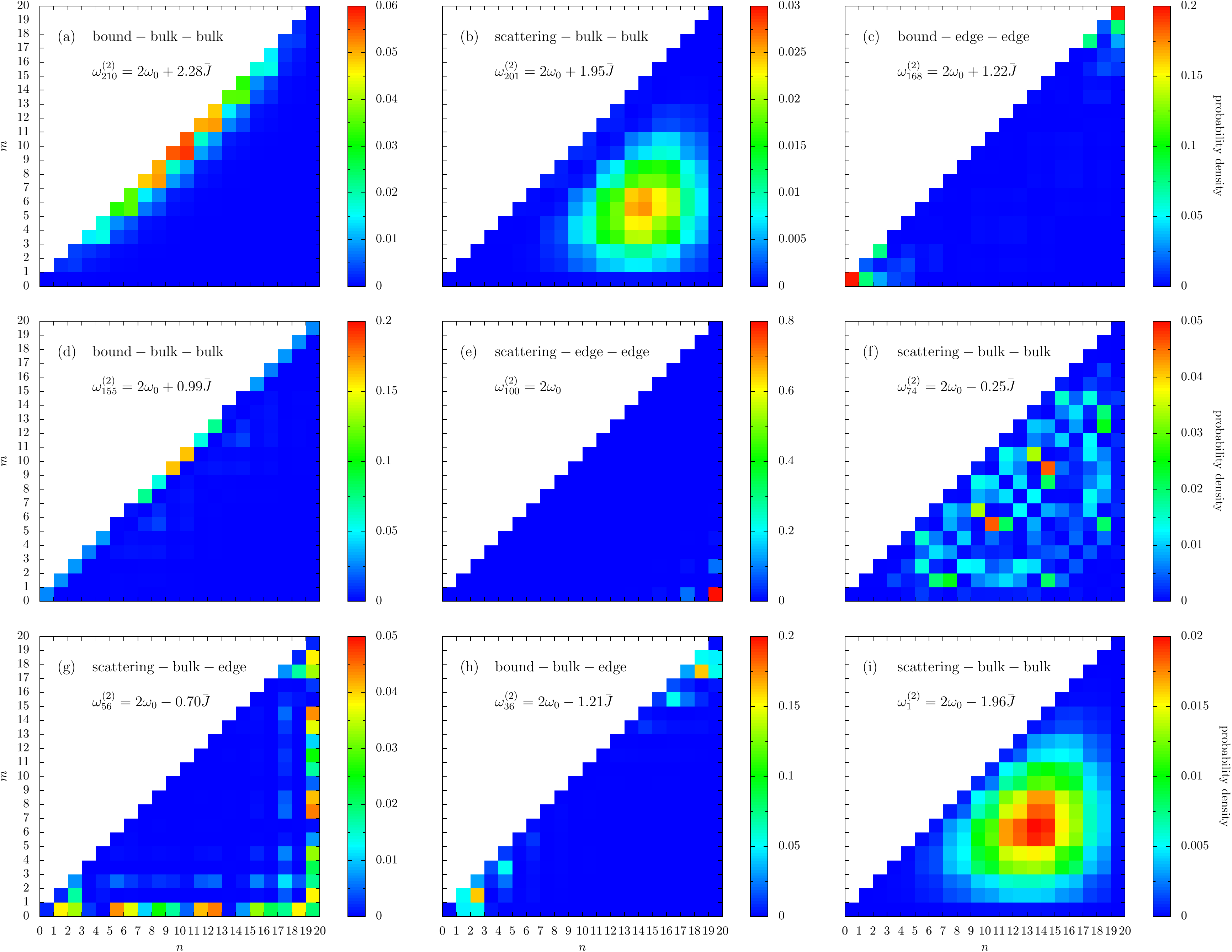}
 \caption{ Nine typical eigenstates of a dimerized chain in the two excitation sector, with nonzero interactions (here, $U=\bar{J}$). Color bars: probability densities (note: the scale is different for each panel). In the figure, the dimerization parameter $\epsilon = -1/2$ and the chain is composed of $N = 10$ dimers, leading to a $210$-dimensional Hilbert space. This regime corresponds to descending the thin, gray vertical line in Fig.~\ref{diagSSH}~(c).}
 \label{stateUone}
\end{figure*}

In Fig.~\ref{stateUone} we show the impact of interactions on the system at the level of the eigenstates, similar to Fig.~\ref{stateUzero} but now with nonzero interactions. In particular, in Fig.~\ref{stateUone} we concentrate on the case of $U = \bar{J}$ and $\epsilon = -1/2$, corresponding to the solid gray vertical line in panel (c) of Fig.~\ref{diagSSH}. As we move through the panels (a-i) in Fig.~\ref{stateUone}, we look at states with increasingly small eigenfrequencies $\omega_{m}^{(2)}$ (that is, we go from top to bottom down the vertical line in panel (c) of Fig.~\ref{diagSSH}). In Fig.~\ref{stateUone}~(a), we show an example of what we call a bound-bulk-bulk state. This nomenclature refers to the state belonging to a bound state band [rather than a scattering band, which are marked by the dashed gray lines in Fig.~\ref{diagSSH}~(c)], and infers that the state is composed of two bulk-like (rather than edge-like) single particle states. Characteristically, the bound-bulk-bulk state result displayed in Fig.~\ref{stateUone}~(a) is of a high probability density on the diagonal (a feature of bound states) which are concentrated in the bulk of the diagonal rather than at the edges. In Fig.~\ref{stateUone}~(b) we have entered a scattering band, and we show an example scattering-bulk-bulk state, with a typical highly spread probability density all over the triangular eigenspace. A signature of topology is found in Fig.~\ref{stateUone}~(c), a so-called bound-edge-edge state (see the red squares in the doublon corners), and a state which does not appear in the geometrically opposite case of a dimerized chain with $\epsilon = +1/2$ [see Fig.~\ref{diagSSH}~(c)]. The state in Fig.~\ref{stateUone}~(d) is a bound-bulk-bulk state, similar to the case of panel~(a), but significantly more doublonic as can be seen by the lack of leakage from the diagonal $n=m$. Exactly at $2 \omega_0$ resides a fully topological scattering-edge-edge state as shown in Fig.~\ref{stateUone}~(e), which is tightly confined to the first and last oscillators in the chain (see the red square in the right-angle corner). Upon further decreasing in frequency, panel (f) returns to a scattering band, and shows an extended scattering-bulk-bulk state with a distinctive checkerboard profile. Slightly lower in frequency is another scattering-bulk-edge state, as displayed in panel (g), with the representative `reflected-L' shape. A bound bulk-edge state is shown in Fig.~\ref{stateUone}~(h), where the two excitations are on the same dimer but not on the same oscillator site (hence it shows significant leakage from the doublon diagonal $n=m$). Finally, panel (i) shows an example state from the lowest band, a scattering-bulk-bulk state with a standard bump-like probability profile similar to panel (b).

Clearly, non-negligible interactions have introduced a rich smorgasbord of possible states in Fig.~\ref{stateUone}, especially in comparison to the simple interactionless case discussed in Fig.~\ref{stateUzero} [and even more so when compared to the single-excitation sector results alluded to by Fig.~\ref{SSHdim}]. The features presented in the eigenfrequency-dimerization plots shown in Fig.~\ref{diagSSH} provide credence to the view that the model of Eq.~\eqref{SSH_Ham_1} exhibits interesting and nontrivial many-body effects already in the two-excitation subspace, including new bands in the bandstructure and novel types of topological states, which may be readily probed in various experimental platforms due to the generality of the theoretical setup.

\section{Discussion}
\label{Conc}

We have studied the two-particle Bose-Hubbard version of the celebrated Su-Schrieffer-Heeger topological model. Our comprehensive survey encompasses the complete range of possible dimerizations of the chain, including both the weakly dimerized and strongly dimerized regimes (where the relevant physics is captured by the Bose-Hubbard dimer). In doing so, we have been able to characterize some general properties of one-dimensional topological lattices with nonzero on-site interactions. Most importantly, the fingerprint of one-particle topology remains in the two-excitation sector, and manifests itself in the creation of new bands in the two-particle bandstructure, and novel types of highly localized state in the topological phase (as defined by the Zak phase in the single-particle sector). We have systematically reviewed the exotic gamut of possible eigenstates supported in the array, which are associated with distinctive patterns in their probability densities and thus are ripe for experimental detection.

On the experimental side, recently there have been a number of exciting reports probing interacting Su-Schrieffer-Heeger-like setups using platforms as diverse as topolectrical circuits~\cite{Olekhno2020} and arrays of superconducting qubits~\cite{Besedin2020}. Meanwhile, interaction effects in related topological models have been realized using optical lattices~\cite{Preiss2017}, Bose-Einstein condensates~\cite{Tai2017}, Rydberg atoms~\cite{Leseleuc2019}, and a 24-qubit superconducting processor~\cite{Ye2019}. Hence, the theoretical results that we have reported should have some utility in a plethora of modern experimental systems with the scope to modulate the strength of interactions in an array of resonators~\cite{Roushan2017b, Roushan2017, Ma2019}.

Our work also opens up several avenues for further study, including: understanding the influence of dissipation~\cite{Naether2015, Quijandria2015, Kordas2015, Lyubarov2018}, exploring unconventional quantum transport effects~\cite{Fedorov2021}, going beyond our strictly enforced two-excitation subspace to probe higher multi-excitation phenomena, and investigating the impact of broken symmetries in the dimer (leading to an interacting, bosonic Rice-Mele-like topological model~\cite{Rice1982}).


\section{Methods}
\label{Metho}

The theoretical results of the main text simply result from diagonalizing the matrix representation of the Hamiltonian of Eq.~\eqref{SSH_Ham_1}. However, the detailed Supplementary Information~\cite{SuppInfo} contains both analytic and numeric results supporting the principal results reported in the main text. The Supplementary Information includes a systematic review of the anharmonic oscillator, the Bose-Hubbard dimer, the Bose-Hubbard chain, and the Bose-Hubbard dimerized chain, and utilizes methods such as bosonic Bogoliubov transformation, Bethe ansatz, and topological band theory. 


\section*{Acknowledgments}
\textit{Funding}: CAD is supported by the Royal Society via a University Research Fellowship (URF\slash R1\slash 201158) and a Royal Society Research Grant (RGS\slash R1 \slash 211220). PAM is grateful to the University of Luxembourg for hospitality. \textit{Data availability}: This study did not generate any new data.  \textit{Code availability}: The code is available at \href{https://github.com/fame64/2ParticleTopologicalSSH}
{https://github.com/fame64/2ParticleTopologicalSSH}, and is curated by P.~Mart\'{i}nez~Azcona (e-mail: pablo.martinez.azcona@gmail.com). \textit{Discussions}: We thank S.~Mart\'{i}nez~Azcona for technical support and D.~Zueco and L.~Mart\'{i}n-Moreno for fruitful discussions.




\begin{thebibliography}{100}
\bibitem{note1}
 More specifically, the scaling with system size $\mathrm{PR}(m) \propto N^\alpha$ exposes both extended states (which typically may be described with $\alpha \ge 1$) and localized states (where usually $\alpha \simeq 0$).
\end{thebibliography}
\end{document}